# Hydrophobic Repulsion and its Origin

Itai Schlesinger and Uri Sivan

*Department of Physics and the Russell Berrie Nanotechnology Institute, Technion - Israel Institute of Technology, Haifa, 32000, Israel*

**The fundamental role of hydrophobic interactions in nature and technology has motivated decades long research aimed at measuring the distance-dependent hydrophobic force and identifying its origin. This quest has nevertheless proved more elusive than anticipated and the nature of the interaction at distances shorter than 2-3 nanometers, or even its sign, have never been conclusively determined. Employing an ultra-high resolution frequency-modulation atomic force microscope (FM-AFM) we succeeded in measuring the interaction at all distances and discovered that the commonly observed attraction at 3-10 nanometer distances turns into pronounced repulsion below 0.3-3 nanometers, depending on the hydrophobicity of the AFM tip and the surface. This generic short-range repulsion disproves the prevailing dogma that hydrophobic interactions are attractive at all distances, hence bearing on practically all biological and technological hydrophobic phenomena. The short-range repulsion, as well as the mid-range attraction, are traced by experiment and theory to the prevalent air accumulation near hydrophobic surfaces.**

## Introduction

The interface between water and hydrophobic surfaces[1] is ubiquitous in nature and technology and so is water mediated interaction between hydrophobic objects. The dominant role of hydrophobic interactions in diverse phenomena such as natural and manmade membranes, proteins, colloids, micelles, biological and manmade ion channels, contaminants in water, gels, chromatography columns, sprays, and more, has spawned extensive research over several decades, aimed at measuring the distance dependent hydrophobic force and identifying the unifying principles underlying it. This quest has nevertheless proved more challenging than anticipated[2-6] and its most important aspect, namely, the nature of the interaction below 2-3 nanometers, and even its sign, have never been conclusively determined.

The interaction between two hydrophobic surfaces immersed in water has been studied extensively by Surface Force Apparatus (SFA) and static Atomic Force Microscopy (SAFM) with or without a colloid glued to the cantilever tip[6-8]. Early experiments reported attractive forces extending over tens and even hundreds of nanometers. Those were found, however, to reflect surface inhomogeneity or the presence of gas bubbles[9]. Water degassing was found to suppress bubble formation and, hence, exposed a shorter attraction range, which nevertheless extended to distances longer than $\approx 10\,\text{nm}$ [10]. Some authors[9] identified this force with separation-induced spinodal cavitation while later measurements found it to be consistent with van der Waals (vdW) attraction[11], as formerly proposed by van Oss *et al.*[12]. It should be noted, though, that vdW attraction is not unique to hydrophobic surfaces. It is seen also with hydrophilic surfaces whenever the long-range double layer interaction is screened out[13]. The "true" hydrophobic interaction, being associated with surface hydration, or lack thereof, is short-ranged and mostly inaccessible to SFA and SAFM due to their inherent mechanical instability in the presence of the large attractive force gradients[6-8] encountered at intermediate distances, 2-5 nanometers. This instability leads to an abrupt jump of the two surfaces to contact and consequently to a blind spot at distances shorter than $\approx 2-5\,\mathrm{nm}$, the most important distance range for many hydrophobic phenomena. In the absence of data it was generally assumed that the attraction observed at intermediate distances persists to shorter distances.

The introduction of liquid frequency modulation atomic force microscopy[14] (FM-AFM) opened a window for probing short-range hydrophobic interactions. Utilizing rigid cantilevers, this dynamic mode of operation was shown to be free of cantilever instability and, hence, capable of probing the rapidly varying hydrophobic interaction at short distances. Using a similar AFM measurement method, Katan and Oosterkamp[15] have measured the interaction between a carbon nanotube tip and a hydrophobic self-assembled alkane-thiol monolayer. In accord with earlier measurements by SFA and SAFM, they found attraction developing at distances shorter than $\approx 5\,\mathrm{nm}$ while at shorter distances, where SFA and SAFM jump to contact, the attractive force grew smaller and eventually turned repulsive below $\approx 1\,\mathrm{nm}$. This measurement of the hydrophobic force at short distances has thus suggested short-range repulsion rather than continuous attraction. At the same time, the inhomogeneous nature of the surfaces used in that work, which comprised nanoscopic hydrophilic and hydrophobic patches, left room for other explanations related to hydration repulsion by the hydrophilic domains.

All in all, despite extensive research, the nature of the short-range hydrophobic interaction, and even its sign, remained unknown. This gap in knowledge motivated the present study.

Using a high-resolution liquid FM-AFM[16] built in-house to measure forces in liquid environment, and combining surfaces and cantilevers of different hydrophobicity levels, we show that the short-range interaction is indeed invariably repulsive, even for homogenous surfaces such as graphite. The range of repulsion grows from $0.2-0.3\,\text{nm}$ in the case of two weakly hydrophobic surfaces to $3\,\text{nm}$ in the case of an extremely hydrophobic surface interacting with a weakly hydrophobic tip.

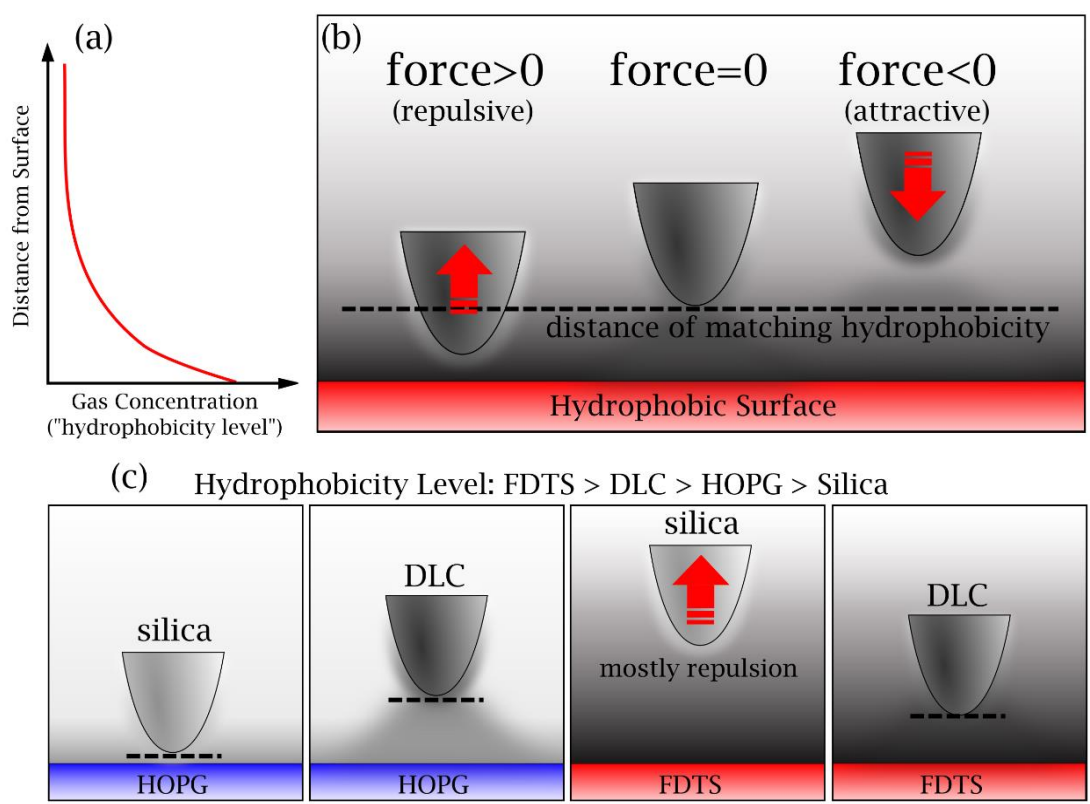

**Fig. 1: Principle of matching hydrophobicity.** Gray gradients above the surface denote hydrophobicity level (average gas concentration) with darker shading designating higher concentration. (a) Air molecules attracted to hydrophobic surfaces create an interfacial layer of graded hydrophobicity level. (b) Objects of intermediate hydrophobicity (i.e., an AFM tip) are driven to a "distance of matching hydrophobicity" where their average interaction with the medium matches the local average interaction between medium molecules. (c) Different combinations of tips and surfaces used in this work as well as their distances of matching hydrophobicity (dashed lines). The FDTS being more hydrophobic than the HOPG surface creates a denser interfacial layer.

The repulsion is traced to the accumulation of air molecules near the hydrophobic surfaces. In equilibrium, the concentration of air molecules near the hydrophobic surface is higher than their concentration in bulk water. Being a continuous function of distance from the surface, the concentration interpolates between these values leading to a few molecule thick interfacial layer of graded density and, hence, hydrophobicity level[1,17] (Fig. 1(a)). When the two hydrophobic surfaces approach each other, their graded atmospheres interact and exert force on the surfaces. The interaction is calculated below within the lattice-gas model but its essential features can be demonstrated in a simplified asymmetric model where the atmosphere created by one of the objects can be neglected. This is the case, for instance, when a small object of intermediate hydrophobicity is placed near an infinite flat hydrophobic surface having a dominant effect on the graded gas-water mixed layer. At larger distances from the surface, where the mixture comprises fewer air molecules, the object, being more hydrophobic than its surroundings, is driven towards the surface while at shorter distances, where the air molecules density is higher and the mixture is more hydrophobic than the object, it is repelled from it (Fig. 1(b)).

The same mechanism hence leads to both medium-range attraction and short-range repulsion, depending on the relative hydrophobicity of the surface and the small object. At the point where the force vanishes, the object's hydrophobicity matches the local medium hydrophobicity, hence the term "distance of matching hydrophobicity" used in the present manuscript.

Note that in the general, more symmetric case, the object itself modifies the local concentration of air in the mixture as seen in Fig. 1 and in the solution of the full lattice-gas model (Fig. 3).

A partial, yet illuminating analogy would be a hot air balloon hovering at an altitude where its average specific mass matches that of the surrounding atmosphere. At lower altitudes the atmosphere is denser and hence pushes the balloon up while at higher altitudes the opposite is true. Similarly to the hydrophobic case, the repulsion of the balloon from the surface of earth at low altitudes is induced by the stronger attraction of dense air to earth. Matching hydrophobicity then maps onto matching average specific mass. The analogy is obviously limited but it gives the flavor of both the longer-range attraction and the short-range repulsion observed in the hydrophobic case. In particular, it is clear that in both cases repulsion requires the presence of air. The proposed mechanism is in a sense Archimedes' principle applied to a strongly interacting system.

## Experimental Results

The experiments were carried out on four tip-surface combinations comprising two surfaces: perfluorodecyltrichlorosilane (FDTS, see supplementary material) monolayer on silicon and highly oriented pyrolytic graphite (HOPG), and two tips: diamond-like carbon (DLC) and silica. The corresponding hydrophobicity levels satisfy FDTS>DLC>HOPG>Silica (Fig. 1(c)).

Fig. 2(a) depicts the force, $F$, versus the tip-surface distance, $h$, acting between a silica or a DLC tip and a silicon wafer coated with FDTS in the presence of deionized water (Supplementary material Fig. S1). A droplet of degassed deionized water was placed on the substrate and the force acting between the cantilever tip, immersed in that droplet, and the surface was measured using FM-AFM. Frequency-shift vs. distance curves[18] were recorded at the times indicated in the legend (Supplementary material

Fig. S2) and converted to force[19]. The water droplet remained exposed to atmosphere throughout the experiment, allowing air diffusion and its spontaneous accumulation at the hydrophobic interface.

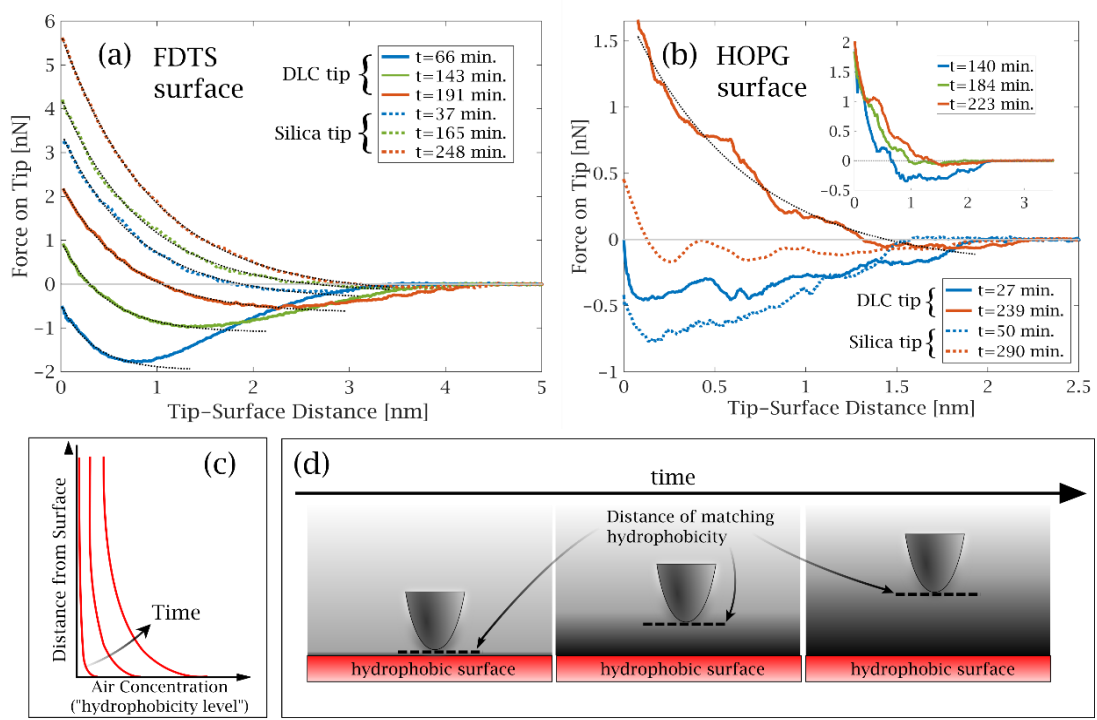

**Fig. 2: Force evolution with time.** (a) Force vs. tip-surface distance measured between FDTS coated substrate and two types of tips in pre-degassed deionized water. Different curves correspond to different exposure times to atmosphere. The dotted black lines depict best exponential fit with the following characteristic lengths: DLC 66 min – 0.34 nm, DLC 143 min – 0.47 nm, DLC 191 min – 0.70 nm, silica 37 min – 0.79 nm, silica 165 min – 0.85 nm, silica 248 min – 0.95 nm. (b) Main panel: Same as (a) for HOPG substrate. Dotted black line depicts best exponential fit with 0.65 nm characteristic length. Inset: Three additional force curves measured with a DLC tip. (c) Evolution of graded hydrophobic layer with time. (d) The corresponding growth of the distance of matching hydrophobicity with time.

As time elapsed, and more air accumulated at the FDTS-water interface, the short-range repulsion grew for both tips (Fig. 2(a)). Experiments with non-degassed water displayed force curves similar to those measured with degassed water after $3-4\,\mathrm{h}$ exposure to atmosphere (data not shown). In accord with the principle of matching hydrophobicity, the overall repulsion was stronger and longer ranged with the less hydrophobic silica tip compared with the more hydrophobic DLC tip.

Repulsion could be fitted with decaying exponents for both tips, at all times. The decay lengths varied between 0.34 nm for the more hydrophobic DLC tip, one hour after immersion, to 0.95 nm in the case of silica tip, four hours after immersion. The decay lengths were generally shorter for the more hydrophobic DLC tip, and in both cases grew larger with accumulation of air near the interface. The force curves were found to be independent of salt concentration, at least up to 100 mM NaCl.

The main panel of Fig. 2(b) depicts the same type of measurements, carried out with the same types of tips against an HOPG substrate, which is less hydrophobic than FDTS (see Fig. S3 for frequency-shift curves). The attraction diminished with time for both tips but in the case of the less hydrophobic silicon tip, repulsion commenced only at very short distances, $0.2 - 0.3$ nm, in agreement with Ref. 20. In the case of the more hydrophobic DLC tip, a full-fledged exponential repulsion evolved, similarly to the case of FDTS substrate shown in Fig. 2(a). More force curves, measured in a different experiment using the latter tip and substrate are shown in the inset to Fig. 2(b) after relatively long exposure to atmosphere.

Note that contrary to the trend in Fig. 2(a), repulsion in the case of HOPG was stronger with the more hydrophobic DLC tip compared with the silica one. This inversion stems from differences in the substrate hydrophobicity and follows the principle of matching hydrophobicity; two objects of comparable hydrophobicity tend to repel each other only feebly compared with objects of markedly different hydrophobicity. FDTS being very hydrophobic (112° contact angle) repels the weakly hydrophobic silica tip stronger than it repels the more hydrophobic DLC tip. Similarly, the mildly hydrophobic graphite (

60° contact angle) repels the weakly hydrophobic silica tip very feebly compared to its repulsion of the moderately hydrophobic DLC tip (Fig. 1(c)).

Finally, note that the HOPG force curves show wiggles spaced by $\approx 0.5-0.6\,\text{nm}$. This modulation shows clearly in the frequency-shift curves depicted in Fig. S3. Its periodicity is considerably longer than the $\approx 0.3\,\text{nm}$ period found with hydrophilic surfaces[21], its magnitude was larger, and it decayed slower with distance. Weak modulation was also observed with FDTS and hydrophobic tip (Fig. S2). Associating the frequency-shift modulation with molecular layering, we find that the attractive interaction and air accumulation are restricted to 3-4 molecular layers.

Integration of the force depicted in Figs. 2 with respect to tip-surface distance, gives the potential of mean force (PMF), or constrained free energy, depicted in Figs. 3(a), and 3(b) for FDTS and HOPG, respectively

$$\Phi(h) = \int_{h}^{\infty} F(z)\,dz \quad .$$

The repulsive and attractive PMFs vary in the range of $3 \div 5 \times 10^{-18}\,\text{J}$. Approximating the relevant tip area by $2\pi a^2$ with $a = 2\,\text{nm}$, and dividing the measured free energy change by this area, one finds free energy density of $0.12 \div 0.20\,\text{J/m}^2$, twice to three times the water surface tension, $0.07\,\text{J/m}^2$. The observed interaction is hence consistent with the hydrogen bond physics of the proposed mechanism rather than with the order of magnitude weaker vdW interaction. To test the potential contribution of electrostatic interactions to the total force the experiments were repeated in a 100mM NaCl solution characterized by a short screening length (1nm). No measurable effect has been observed compared with measurements in pure DI water, indicating that electrostatic effects can also be neglected in the analysis.

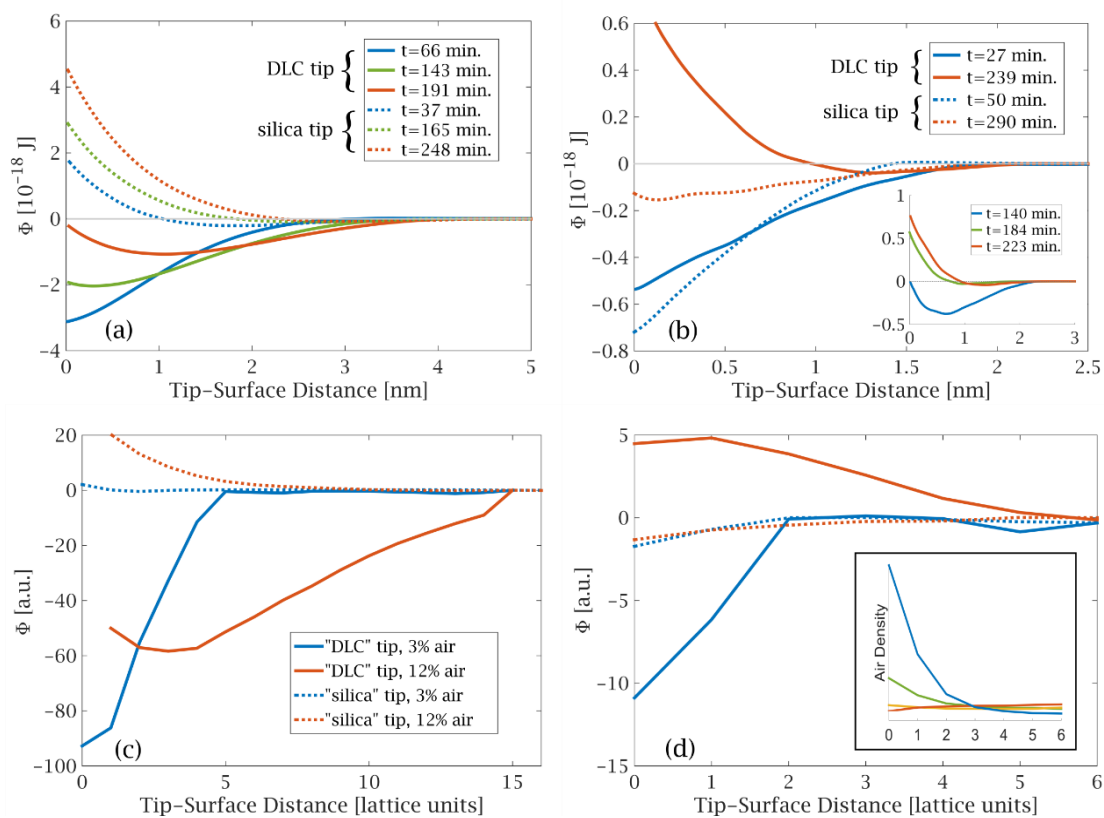

**Fig. 3: Experimental and simulated potentials of mean force.** (a) PMF calculated from the FDTS force curves of Fig. 2(a). (b) PMF calculated from the HOPG force curves of Fig. 2(b) and its inset. (c) PMF calculated by Monte Carlo simulation for the strongly hydrophobic "FDTS" substrate. (d) Same as (c) for the mildly hydrophobic "HOPG" substrate. Inset: Average air density vs. distance for four values of the surface hydrophobicity parameter, $\alpha=0$ (blue), $\alpha=0.2$ (green), $\alpha=0.5$ (gold), $\alpha=1$ (red). Average air concentration in simulation box was 5%.

The PMFs for HOPG, depicted in Fig. 3(b), are significantly smaller in magnitude than those measured with FDTS (Fig. 3(a)), in accord with the mild hydrophobicity of graphite. In fact, the interaction with silica, another weak hydrophobe, is mostly attractive, except for long exposure times at very short distances.

Simulations

The principle of matching hydrophobicity is readily formulated for immiscible fluids. Consider, for instance, the interface between water and oil. Immiscibility implies, by definition, that a droplet of oil (or comparable hydrophobic colloid) put in water and a droplet of water (or comparable hydrophilic colloid) put in oil will find themselves for finite size samples back in their native medium where their hydrophobicity level matches the medium.

In the case of partially miscible fluids, such as air and water, the situation is complicated since the solute molecules are mobile and rearrange themselves when the tip approaches the surface. To simulate this situation we resort to Monte Carlo simulations of a three component lattice-gas model on a square two-dimensional lattice with periodic boundary conditions. Air molecules are emulated by $\uparrow$ spins, water by $\downarrow$ spins, and the tip and surface by a third type of spin, which is a (classical) linear combination of $\uparrow$ and $\downarrow$ spins, i.e., $\updownarrow = \alpha \downarrow + (1-\alpha) \uparrow$. The nature of the $\updownarrow$ spin can be tuned continuously between air and water when the parameter $\alpha$ is varied between zero and one. The corresponding nearest neighbor interaction energies (measured in $k_B T$ units) for air-air, air-water, water-water, $\updownarrow$-air, and $\updownarrow$-water are $0, J, -J, \alpha J, (1-2\alpha)J$, respectively, where $J$ is a positive parameter. The surface is emulated by a line of $\updownarrow$ spins, spanning one side of the simulation square. The tip is emulated by a line of $\updownarrow$ spins, usually of a different $\alpha$ than the surface. Tip movement is confined to a straight line, perpendicular to the surface. The canonical micro-states space is probed using the Metropolis-Hastings algorithm[22]. A histogram of tip position, $P(h)$, is generated over $3 \times 10^7$ simulation steps after the system has reached equilibrium and the potential of mean force acting on the tip is calculated, $\Phi(h) = -\ln(P(h))$.

The simulations presented in Figs. 3(c,d) were conducted on a $50 \times 50$ sites lattice with the surface emulated by a $50$ sites line, the medium emulated by $2440$ "molecules", and the tip emulated by a $10$ sites long line parallel to the surface. The number of $\uparrow$ spins was kept at either 3% or 12% of the medium molecules number to emulate the experimental accumulation of air molecules at two different times. The coupling constant was set to $J = 2/3$. The hierarchy of experimental hydrophobicity, $\text{FDTS} > \text{DLC} > \text{HOPG} > \text{silica}$, was mimicked by $\alpha = 0, 0.2, 0.8, 0.9$, respectively.

Despite the crude nature of the model it reproduces the main experimental features remarkably well. Those include attraction, repulsion, and the cross-over between the two. Fig. 3(c) depicts the tip PMF for an emulated FDTS surface and should be compared with the corresponding experimental results depicted in Fig. 3(a). The two concentrations of air molecules mimic measurements after short (3% air) and long (12% air) exposures to atmosphere. Similarly to the experiment, the simulated PMFs for the less hydrophobic "silica" tip disclose short-range repulsion growing larger with air concentration. In the case of the more hydrophobic "DLC" tip we find long-range attraction, starting farther from the surface at the higher air concentration. In accord with the experimental results, attraction persists to contact in the case of low air concentration while at a higher air concentration, a short-range repulsive component appears. As predicted by the principle of matching hydrophobicity, the distance of minimal PMF corresponds to the point where the average tip interaction with the medium equals the average interaction between medium molecules. The missing point at zero distance in the 12% "DLC" tip reflects strong repulsion from the surface; the tip did not spend any time there in $3 \times 10^7$ simulation steps.

Fig. 3(d) depicts the tip PMF for an emulated HOPG surface and two concentrations of air molecules. Now, the surface hydrophobicity is slightly higher than "silica" and smaller than "DLC". Similarly to the experiment, the interaction at low air concentration (short times) is attractive for both tips while for high air concentration it is repulsive for the more hydrophobic "DLC" tip and attractive in the case of "silica" tip. This result concurs again with the principle of matching hydrophobicity and the experimental results. The inset to Fig. 3(d) depicts air concentration vs. distance for different surfaces, in the absence of a tip. The graded concentration which leads to the short-range repulsion and mid-range attraction is clearly seen with the more hydrophobic surfaces.

The principle of matching hydrophobicity takes a simple form here. The spins emulating the tip reflect linear combination of air and water. At the distance of minimal PMF (zero force), the relative weight of the two components matches the relative average local concentration of air and water and the free energy is hence inert to its presence. Moving the tip either closer or farther from the surface generates spin configuration which is different from the equilibrium one and, hence, less favorable.

Discussion

Dissolved gases such as nitrogen, oxygen, and carbon dioxide are abundant in most biological and technological systems. In fact, special measures should be taken to remove those gases from solution in cases where they play a detrimental effect. The effect of dissolved gas on water depletion next to hydrophobic surfaces was observed in neutron scattering[23-25] and in numerical simulations[26,27]. The effect of dissolved gas on bubble formation and spontaneous cavitation, and the resulting long-range attraction, were studied by SAFM[8-10,28,29] (and references therein). Finally, the formation of gas-rich stripes on HOPG exposed to gas-saturated solution has been observed by FM-AFM[30,31]. We are not aware of previous reports on the gradual suppression of short-range attraction by accumulating gas molecules and the eventual replacement of attraction by pronounced repulsion, probably due to the blind spot of the widely used SFA and SAFM at nanometer distances. The recent introduction of force spectroscopy by liquid FM-AFM overcame this hurdle and facilitated the present discovery.

The overriding role of gas molecules reported here, necessitates reconsideration of most theories involving hydrophobic interactions since with few exceptions (e.g., Refs. 25-27), those did not take into consideration dissolved gases[32]. The discovered short-range repulsion should modify, for example, models of the interaction between alkane chains or proteins, whose self-assembly leads to biological and manmade membranes. Equally important, the proven role of accumulated gases will certainly affect the analysis of data collected by FM-AFM, neutron, and x-ray scattering.

Regarding the thickness of the gas-rich interfacial layer, Lum, Chandler and Weeks[33] have shown theoretically that water should be depleted next to macroscopic hydrophobic surfaces, regardless of air accumulation. Extensive theoretical and numerical work has indicated that such depletion should be limited to about $0.2$-$0.4\,\text{nm}$ [4,5,34,35] from the surface. These results were found to be qualitatively consistent with x-ray reflectance experiments[34-37]. A later experiment[38] found that the higher the hydrophobicity the thicker the depletion layer, reaching $0.6-0.8\,\text{nm}$ for a $120^0$ contact angle. Neutron reflectivity experiments reported conflicting results. No depletion layer was found in the case of deuterated polystyrene films[39] while in the case of functionalized self-assembled monolayers one experiment[5,40] reported depletion lengths similar to those found by x-ray reflectivity and two other experiments disclosed water depletion up to $4\,\text{nm}$ [23] and $2\text{-}5\,\text{nm}$ [24] from the surface. In the former case, gas accumulation near the interface was invoked in order to explain the unexpectedly large depletion length while in the latter case the depletion length was found to depend on the level of air saturation of the water sample and on the time elapsed from contacting it with the hydrophobic surface.

Our measurements found layer thicknesses varying between $0.2-0.3\,\text{nm}$ for the least hydrophobic pair comprising HOPG surface and silica tip at short times, to $1.5\,\text{nm}$ in the case of HOPG-DLC and $3\,\text{nm}$ for the most hydrophobic surface, FDTS, in combination with the least hydrophobic silica tip. The latter thickness was comparable to those found by neutron scattering[23,24].

## Methods

Fresh HOPG surfaces (SPI Supplies) were exposed prior to each experiment by peeling off graphite layers using an adhesive tape. FDTS monolayers were deposited in the gas phase (MVD100E, Applied Microstructures) on silicon (100) wafers. The wafers were cleaned, prior to deposition, with 3:1 $N_2SO_4:H_2O_2$ solution for 15 minutes and then oxidized with $O_2$ plasma (100W, 60 sec., 200mTorr) inside the MVD100E chamber. Experiments were carried out in 18MΩ×cm deionized (DI) water degassed by vacuum pumping over several hours (membrane pump). Silicon tips (ppp-NCH-AuD, Nanosensors) were pretreated with 100W oxygen plasma for 60 seconds (MVD100E, Applied Microstructures) while the DLC tips (MSS-NCH-AuD, Nanotools) were pretreated in UV-ozone to remove organic contaminants.

Contact angles of DI water were measured using the sessile drop technique in Ramé Hart goniometer (Model 250). The contact angle on a freshly exposed HOPG surface was measured to be $60°\pm1°$. The advancing (receding) contact angle on a freshly prepared FDTS sample was $112°$ ($109°$) $\pm 1°$. Due to the small size of the tips it was impossible to measure their contact angle. The contact angle of DLC was reported in Ref. 41 to be in the range of $74°-88°$, namely higher than for HOPG and in agreement with the hierarchy of hydrophobicity levels concluded from the experimental force measurements. To model the silica tip we have measured the contact angle of a flat silicon (100) wafer treated the same way as the silicon tip. The observed contact angle, $3°\pm 1°$, was small but finite (non-wetting), proving that the interaction of water with silica is indeed weaker than water-water interaction.

AFM measurements were carried out using an ultra-high resolution AFM built in-house[16]. The cantilever oscillation amplitude varied between $0.1-0.3$ nm where the results were independent of it.

## Acknowledgement

Research was supported by the Israeli Science Foundation through grant number 10/1051 and the single molecule I-Core center of excellence, grant number 1902/12. We are grateful to Roland Netz, Amos Ori, Joseph Avron, Daniel Podolsky, and Netanel Lindner for numerous discussions.

Corresponding author: phsivan@tx.technion.ac.il

# Hydrophobic Repulsion and its Origin

## Supplementary Material

Itai Schlesinger and Uri Sivan*

*Department of Physics and the Russell Berrie Nanotechnology Institute, Technion - Israel Institute of Technology, Haifa, 32000, Israel*

*To whom correspondence should be sent - phsivan@tx.technion.ac.il

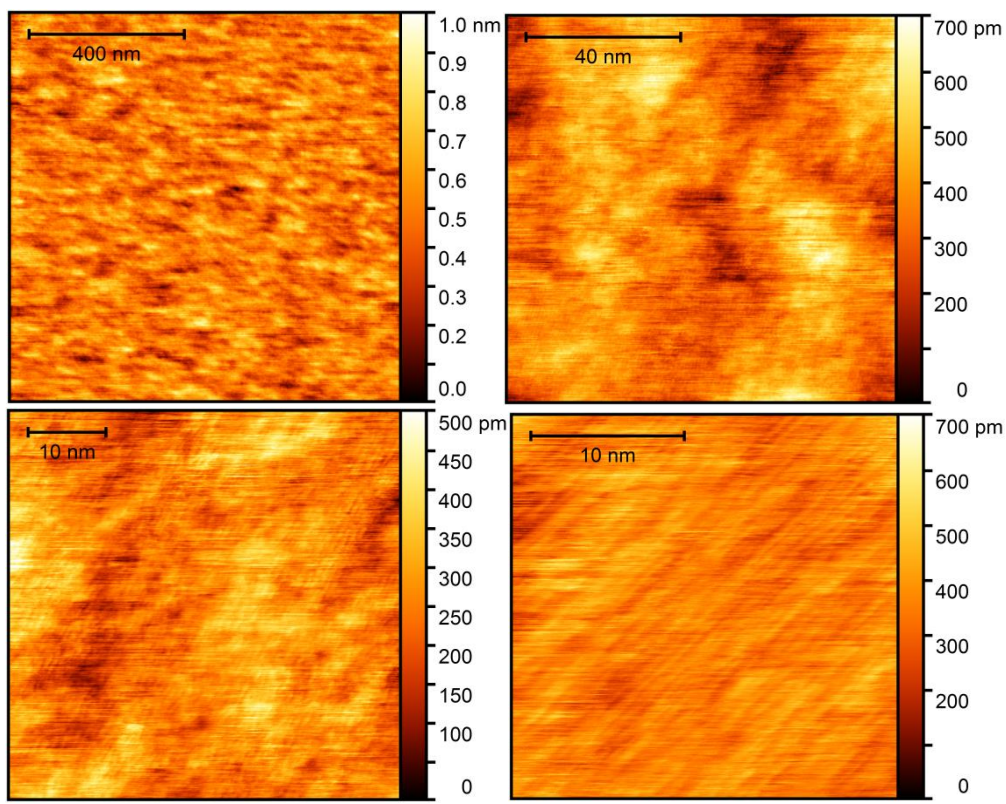

**Fig. S1: FM-AFM images of FDTS monolayer**. FM-AFM images of FDTS monolayer on silicon at increasing levels of magnification. The height variation measured over the 1x1 micrometer field was smaller than <1nm peak to peak and that measured over the 25x25 nanometer field was smaller than 0.4 nm. The stripe pattern observed in high magnification discloses the row-like assembly of the FDTS molecules on the underlying silica surface.

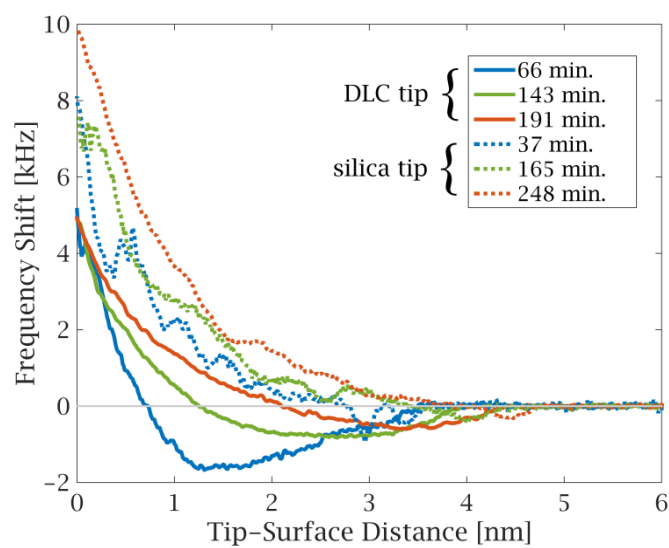

**Fig. S2: FDTS frequency shift vs. tip-surface distance curves.** Raw data collected over an FDTS surface with the two types of tips mentioned in the legend.

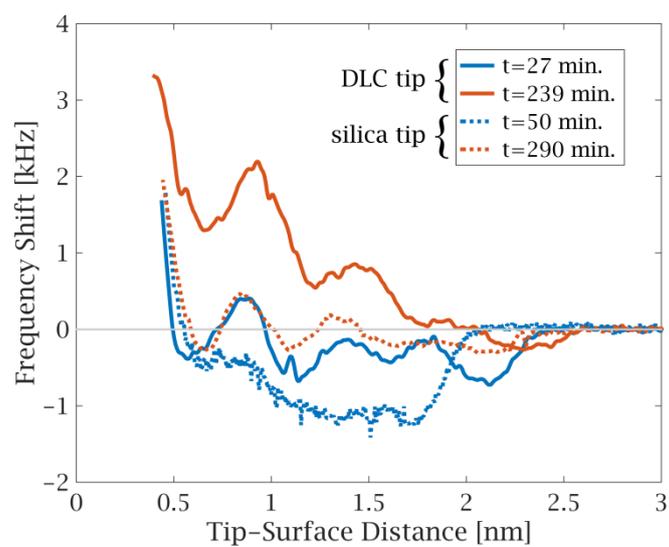

**Fig. S3: HOPG frequency shift vs. tip-surface distance curves.** Raw data collected over an HOPG surface with the two types of tips mentioned in the legend. Note the 0.5-0.6 nm period modulation attributed to molecular layering of the air-gas mixture on the HOPG surface.